\documentclass[conference]{IEEEtran}
\IEEEoverridecommandlockouts

\usepackage[utf8]{inputenc}
\usepackage[T1]{fontenc}
\usepackage{url}
\usepackage{graphicx}
\usepackage{array}
\usepackage{multirow}
\usepackage{verbatim}
\usepackage{hyphenat}
\usepackage{float}
\usepackage{capt-of} 
\usepackage{tcolorbox}
\usepackage{enumitem}
\usepackage{balance}
\usepackage{todonotes}
\usepackage{threeparttable}
\usepackage{colortbl}
\usepackage{epstopdf}
\usepackage{footmisc}
\usepackage{capt-of}
\usepackage{mwe} 
\usepackage{pifont}
\usepackage{mathabx}
\usepackage{cleveref}
\usepackage{tikz}
\usepackage{arydshln}
\usepackage{flushend}

\usepackage{soul}
\setuldepth{abc}

\usepackage{ragged2e}
\usepackage{tabularx}

\newcommand{\etal}{\emph{et~al.}\xspace}

\definecolor{gray50}{gray}{.5}
\definecolor{gray40}{gray}{.6}
\definecolor{gray30}{gray}{.7}
\definecolor{gray20}{gray}{.8}
\definecolor{gray15}{gray}{.85}
\definecolor{gray10}{gray}{.9}
\definecolor{gray05}{gray}{.95}

\definecolor{arsenic}{rgb}{0.23, 0.27, 0.29}

\newtcolorbox{summarybox1}{
	colback=gray!5,
	colframe=arsenic,
	boxrule=0.4mm,
	left=1.5mm,
	right=1.5mm,
	top=1.5mm,
	bottom=1.5mm,
	fonttitle=\bfseries,
	title={Main findings for \textbf{RQ$_1$}}
}

\newtcolorbox{summarybox2}{
	colback=gray!5,
	colframe=arsenic,
	boxrule=0.4mm,
	left=1.5mm,
	right=1.5mm,
	top=1.5mm,
	bottom=1.5mm,
	fonttitle=\bfseries,
	title={Main findings for \textbf{RQ$_2$}}
}

\newtcolorbox{summarybox3}{
	colback=gray!5,
	colframe=arsenic,
	boxrule=0.4mm,
	left=1.5mm,
	right=1.5mm,
	top=1.5mm,
	bottom=1.5mm,
	fonttitle=\bfseries,
	title={Main findings for \textbf{RQ$_3$}}
}

\newlength\Linewidth
\def\findlength{\setlength\Linewidth\linewidth
    \addtolength\Linewidth{-4\fboxrule}
    \addtolength\Linewidth{-3\fboxsep}
}

\newenvironment{rqbox}{\par\begingroup
	\setlength{\fboxsep}{5pt}\findlength
	\setbox0=\vbox\bgroup\noindent
	\hsize=0.95\linewidth
	\begin{minipage}{0.95\linewidth}\normalsize}
	{\end{minipage}\egroup
	\textcolor{gray20}{\fboxsep1.1pt\fbox
		{\fboxsep5pt\colorbox{gray05}{\normalcolor\box0}}}
	\endgroup\par\noindent
	\normalcolor\ignorespacesafterend}

{
}

\newcolumntype{L}[1]{>{\raggedright\let\newline\\\arraybackslash\hspace{0pt}}m{#1}}
\newcolumntype{C}[1]{>{\centering\let\newline\\\arraybackslash\hspace{0pt}}m{#1}}
\newcolumntype{R}[1]{>{\raggedleft\let\newline\\\arraybackslash\hspace{0pt}}m{#1}}

\newcommand{\quickwordcount}{%
  \immediate\write18{texcount -1 -sum -merge \jobname.tex > \jobname-words.sum }%
  \input{\jobname-words.sum} words%
}

\usepackage{hanging}
\newcommand{\ff}[2]{
    \hangpara{0.4em}{1}%
   \makebox[0.4em][l]{\textsuperscript{#1}}{#2}\par%
}

\begin{document}

\definecolor{babyblue}{rgb}{0.63, 0.79, 0.95}
\definecolor{bubblegum}{rgb}{0.99, 0.76, 0.8}
\definecolor{aliceblue}{rgb}{0.94, 0.97, 1.0}
\definecolor{lightsalmon}{rgb}{1.0, 0.63, 0.48}
\definecolor{emerald}{rgb}{0.31, 0.78, 0.47}
\definecolor{grannysmithapple}{rgb}{0.66, 0.89, 0.63}
\definecolor{moccasin}{rgb}{0.98, 0.92, 0.84}
\definecolor{naplesyellow}{rgb}{0.98, 0.85, 0.37}
\definecolor{lightcyan}{rgb}{0.88, 1.0, 1.0}
\definecolor{lightgoldenrodyellow}{rgb}{0.98, 0.98, 0.82}
\definecolor{darktangerine}{rgb}{1.0, 0.66, 0.07}
\definecolor{airforceblue}{rgb}{0.36, 0.54, 0.66}
\definecolor{bostonuniversityred}{rgb}{0.8, 0.0, 0.0}
\definecolor{darkspringgreen}{rgb}{0.09, 0.45, 0.27}
\definecolor{green(ryb)}{rgb}{0.4, 0.69, 0.2}
\definecolor{bittersweet}{rgb}{1.0, 0.44, 0.37}
\definecolor{deepmagenta}{rgb}{0.8, 0.0, 0.8}
\newcommand{\dario}[1]{\todo[inline, color=lightgoldenrodyellow]{Dario: #1}}
\newcommand{\davide}[1]{\todo[inline, color=lightcyan]{Davide: #1}}
\newcommand{\valentina}[1]{\todo[inline, color=bubblegum]{Valentina: #1}}
\newcommand{\rafael}[1]{\todo[inline, color=grannysmithapple]{Rafael: #1}}
\newcommand{\luca}[1]{\todo[inline, color=aliceblue]{Luca: #1}}
\newcommand{\andrea}[1]{\todo[inline, color=moccasin]{Andrea: #1}}
\newcommand{\general}[1]{\todo[inline, color=darkspringgreen]{#1}}
\newcommand{\bline}[0]{\par\noindent\rule{\linewidth}{0.4pt}}

\newcommand{\reviewerOne}[1]{\textcolor{black}{#1}}
\newcommand{\reviewerTwo}[1]{\textcolor{black}{#1}}
\newcommand{\reviewerThree}[1]{\textcolor{black}{#1}}
\newcommand{\reviewerFour}[1]{\textcolor{black}{#1}}
\newcommand{\generalComment}[1]{\textcolor{black}{#1}}
\title{Breaks and Code Quality: Investigating the Impact of Forgetting on Software Development \\ 
{\large Registered Report}}


\author{ \fontsize{10.6pt}{0}\selectfont Dario Amoroso d'Aragona$^1$, Luca Pascarella$^2$, Andrea Janes$^3$,  Valentina Lenarduzzi$^4$, Rafael Pe\~naloza$^5$, Davide Taibi$^1$$^,$$^4$ \\
$^1$Tampere University  --- $^2$ETH Zurich --- $^3$FHV Vorarlberg University of Applied Sciences \\ $^4$University of Oulu --- $^5$University of Milano-Bicocca
\\
dario.amorosodaragona@tuni.fi; lpascarella@ethz.ch; andrea.janes@fhv.at; \\valentina.lenarduzzi@oulu.fi; rafael.penaloza@unimib.it, davide.taibi@oulu.fi 
}

\renewcommand\tabularxcolumn[1]{m{#1}}

\maketitle

\begin{abstract}

Developers interrupting their participation in a project might slowly forget critical information about the code, such as its intended purpose, structure, the impact of external dependencies, and the approach used for implementation. Forgetting the implementation details can have detrimental effects on software maintenance, comprehension, knowledge sharing, and developer productivity, resulting in bugs, and other issues that can negatively influence the software development process. Therefore, it is crucial to  ensure that developers have a clear understanding of the codebase and can work efficiently and effectively even after long interruptions. \reviewerTwo{ This registered report proposes an empirical study aimed at investigating the impact of  the  developer's activity breaks duration and different code quality properties. In particular, we aim at  understanding if the amount of activity in a project impact the code quality, and if developers with different activity profiles show different impacts  on code quality. }
The results might be useful to understand if it is beneficial to promote the practice of developing multiple projects in parallel, or if it is more beneficial to reduce the number of projects each developer contributes.
\end{abstract}

\raggedbottom
\begin{IEEEkeywords} 
Forgetting curve, Code Quality, Empirical Software Engineering
\end{IEEEkeywords}

\section{Introduction}
\label{sec:Intro}

When developers are not working for a long time on the same project, they might forget some details about the source code, including the purpose of some lines of code, the code structure, the effect of external dependencies, or the followed implementation strategy. The result, can can hinder software maintenance, comprehension,  and  developer productivity~\cite{robbes2019}, with possible consequences on bugs, and other issues in software development~\cite{Kruger2018,Dogan2022}. 

\begin{figure}[ht]
    \centering
    \includegraphics[width=0.95\columnwidth]{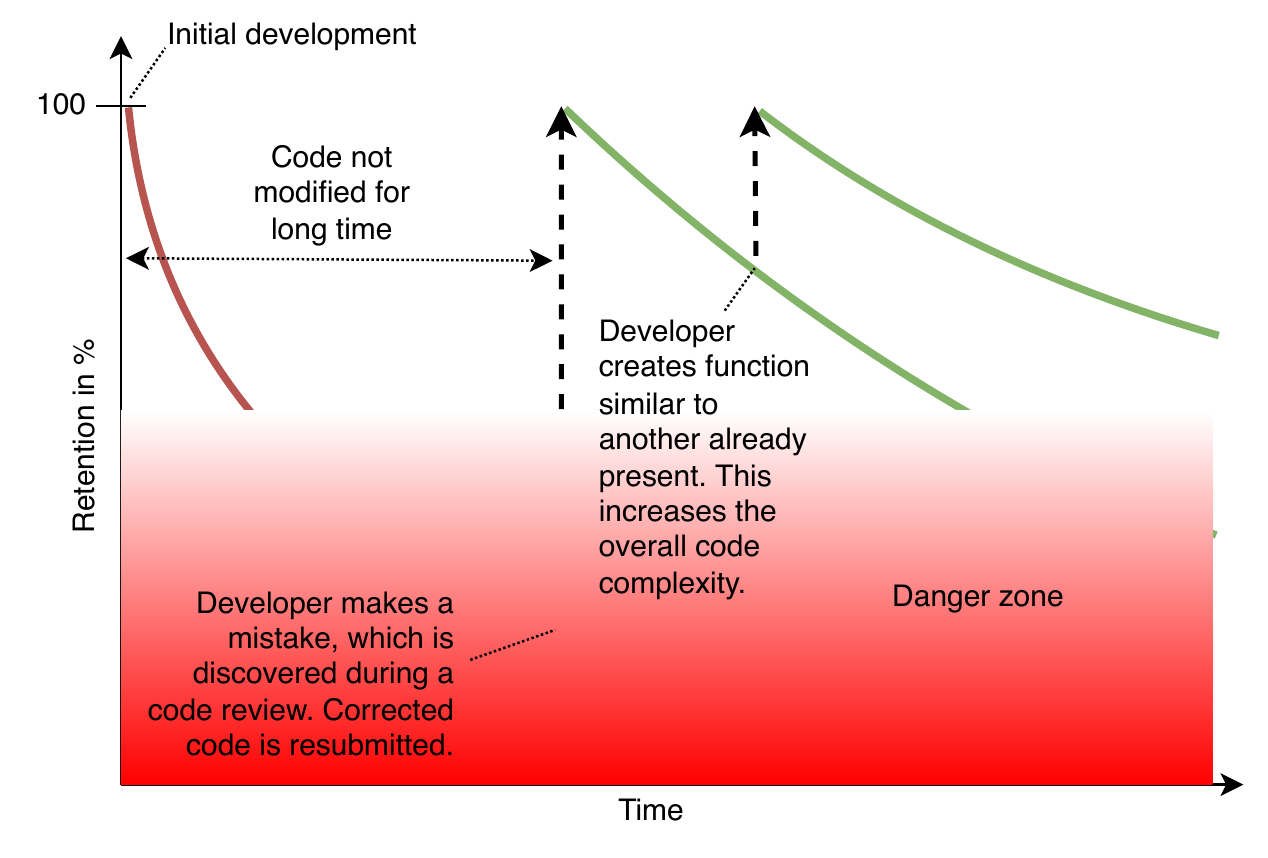}
    \caption{Illustration of the forgetting curve \cite{ebbinghaus1885}, extended with examples that exemplify its application to software development.}
    \label{fig:EbbinghausHypothesis}
\end{figure}

The Ebbinghaus curve is a well-known model for describing a) forgetting as a function of time and b) retaining as a function of repeated learning~\cite{ebbinghaus1885,ebbinghaus2013memory}. Applied to software development (see Figure~\ref{fig:EbbinghausHypothesis}), we hypothesize that when time elapses and a developer does not repeatedly work on a project, he or she might forget some details and might be more prone to introducing mistakes. This was also observed by~\cite{Fritz}, where they reported that ``several subjects each noted that he or she has to work with code [continuously] otherwise I forget after a while [1 month]''. 

Many studies analyze the activity of \textit{learning}. However, the countermeasure to forgetting is not learning but \textit{remembering}. Learning and remembering require different strategies, as re-learning from scratch (to remember) may be considered an inefficient effort and perceived as boring. 

To investigate the phenomenon of code forgetting in more detail and to be able to develop countermeasures, in this registered report, we want to study whether we can observe a relationship between \ul{interruptions during participation in a project} (assuming that these interruptions cause forgetting) and a \ul{degradation of source code quality}. \reviewerThree{Concretely, we operationalize participation as the ``observable, performed activities on the source code repository'' (e.g., commits, pull requests, etc.), interruptions as ``the time that occurred between one activity and the next, performed by the same developer in the same project'', and degradation of source code quality as ``a worsening change in source code metrics''.}

\textbf{Paper Structure:} 
Section~\ref{sec:CaseStudy} describes the empirical study design and Sect.~\ref{sec:Methodology} presents the data collection and analysis protocols. Section~\ref{sec:Execution} outlines the execution plan and Sect.~\ref{sec:ThreatsValidity} identifies the threats to validity. Section~\ref{sec:RelatedWork} identifies the threats to validity, and Sect.~\ref{sec:Conclusion} concludes the paper. 
\section{Empirical Study Design}
\label{sec:CaseStudy}
\begin{figure}[ht]
    \centering
    \includegraphics[width=0.9\columnwidth]{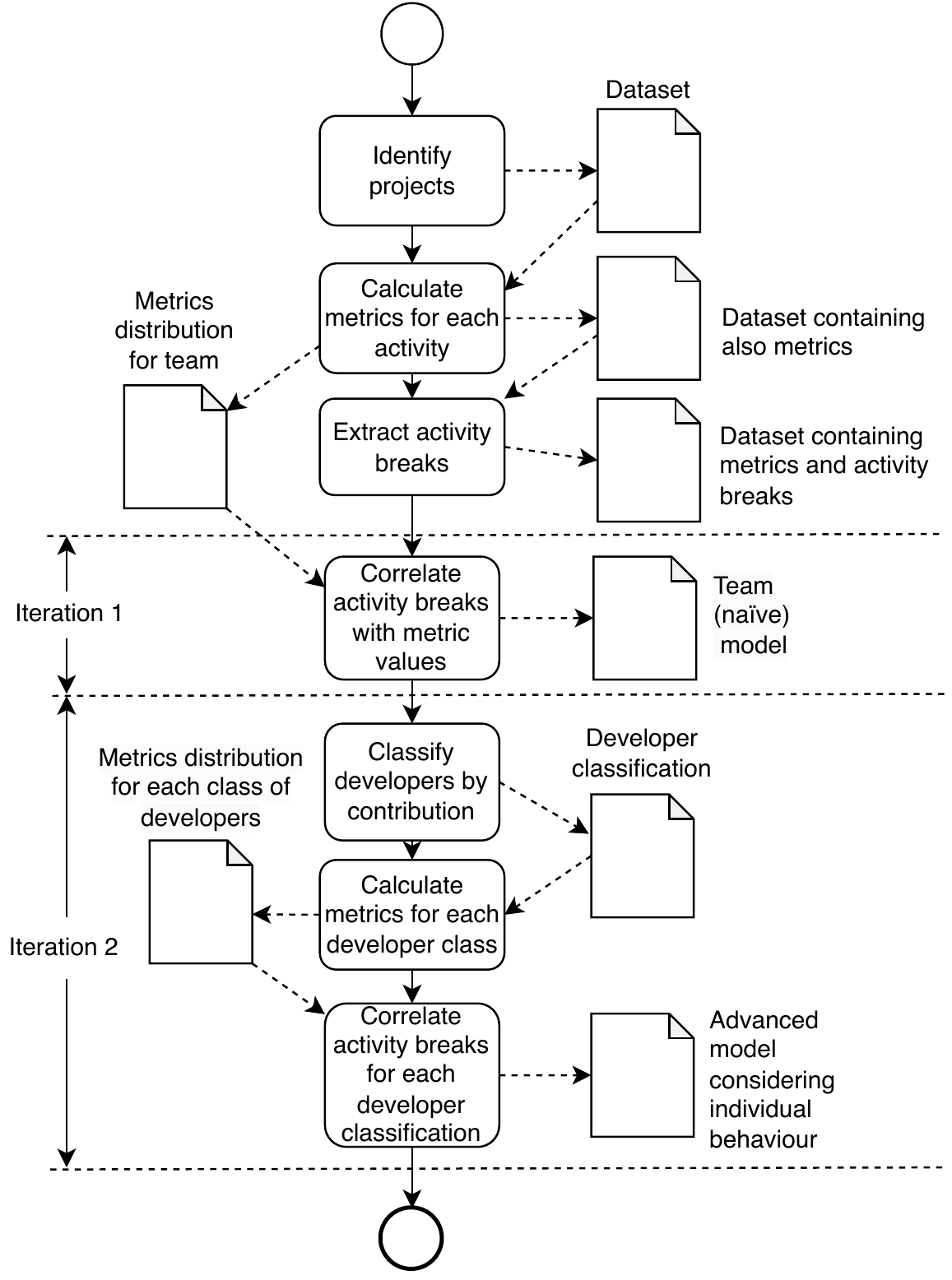}
    \caption{Empirical Study Design Process}
    \label{fig:process}
\end{figure}

In this section, we describe our empirical study reporting the goal and research questions, the context, data collection, and data analysis. We designed our study based on the guidelines defined by Wohlin et al.~\cite{Wohlin2000}. In Figure~\ref{fig:process}, we describe the entire process we will adopt to answer our RQs. 

We split our investigation into two different approaches, hereinafter called ``Iterations''. In Iteration 1, we aim at understanding whether, for a developer in general, the time that elapses between his/her activities correlates negatively with the code quality of the new contribution \reviewerFour{(considered at project and also at module level)} when the developer gets back to the code (since it ignores the personal characteristics of the individual developer, we call this the Naïve model). 

In Iteration 2, we will study if the relationship between interruptions and source code quality degradation can be better explained if the degree of contribution of a developer is also taken into consideration (Advanced model). We assume that primary contributors (authored more than 50\% of the code \cite{TruckFactors}) forget at a slower pace than secondary contributors.

\subsection{Goal, Research Questions, Metrics, and Hypothesis}
\label{sec:Goal}


\reviewerTwo{We formalized the goal of this study according to the GQM approach~\cite{Basili1994}} as \reviewerTwo{\textit{Investigate} interruptions of development activities \textit{for the purpose of} evaluation \textit{with respect to the} impact of their length on source code quality \textit{from the point of view of} developers \textit{in the context of} open-source software}.

To measure source code quality, we will consider readability and quality metrics, see Sect.~\ref{sec:ReadabilityCheckS1}.




Based on the aforementioned goal, we defined two Research Questions (RQ).



\begin{center}	
	\begin{rqbox}
		\textbf{RQ$_1$.} \emph{How strong is the developer activity break duration correlated with a degradation of code quality metrics?}
  \end{rqbox}	 
\end{center}

\reviewerFour{
As ``activity break'' we will consider the time that occurred between the previous activity in the project and the next activity performed by the same developer in the same package.
We consider all activities, which we are able to measure and where we assume that knowledge of the code is required: 
\begin{itemize}
    \item Commits
    \item Opening/closing/reviewing/commenting pull requests 
    \item Opening/closing/commenting issues 
\end{itemize}}

We will collect different \textbf{metrics}. 

\textit{Readability Metrics}. We will measure the readability by using the eight readability metrics  defined by Scalabrino et al.~\cite{scalabrino2018ACM}, which are based on textual properties of the source code, described in Table~\ref{tab:Metrics4}.
Several studies highlighted that textual features are significant descriptors in the evaluation of code comprehension and, therefore, are meaningful indicators of the overall readability level of source code \cite{Latifa2013comprehension, Peitek2021comprehension, Sellitto2022}.
Moreover, Scalabrino et al.~\cite{scalabrino2018ACM} demonstrated that their newly-defined metrics are indeed a proxy of the actual readability perceived by developers.
In other words, the considered metrics are suitable to quantitatively assess the readability of source code and are qualitatively perceived as relevant by practitioners.

\textit{Anti-Patterns and Code Smells}. We will consider the Code Smells defined by Fowler~\cite{Fowler1999} and the anti-patterns defined by Brown~\cite{BrownAntipatterns}
~(\Cref{tab:Metrics1}).

\textit{Software Metrics and Technical Debt detected by SonarQube}. 
We will include software metrics computed by SonarQube as well as the information related to the Technical Debt. SonarQube includes the three categories of issues (Code Smells, Bugs, and Security Vulnerabilities) and the three Technical Debt types (Squale Index, Reliability Remediation Effort, and Security Remediation Effort). We must notice that the Code Smells detected by SonarQube are not the ones defined by Fowler~\cite{Fowler1999}~(\Cref{tab:Metrics2}).
\color{black}

\reviewerTwo{We \textbf{hypothesize} that the antipatterns, code smells, and SonarQube metrics (\Cref{tab:Metrics1} and \Cref{tab:Metrics2}) are directly related to activity break duration (H$_{1.1}$). 
Instead, the readability metrics (\Cref{tab:Metrics4}) are (mostly) inversely related to activity break duration (H$_{1.2}$). In \Cref{tab:Metrics1}, \Cref{tab:Metrics2}, and \Cref{tab:Metrics4}, we report if we expect an increase or a decrease of the relative metric in the rightmost column.}


\begin{center}	
	\begin{rqbox}
		\textbf{RQ$_2$.} \reviewerTwo{\emph{How strong is the developer activity break duration
correlated with the degradation of code quality metrics for classes of developers created according to their participation to a given project?}}
  \end{rqbox}	 
\end{center}
\reviewerTwo{In this RQ we aim at understanding if developers with similar activity profiles (e.g. the super active,
active, average, inactive, and super inactive) have a different impact on code quality.} 

\reviewerTwo{As for \textbf{metrics}, we will consider the same ones adopted for RQ$_1$ but applied to clusters of developers with similar activity profiles.}
\reviewerOne{To cluster the developers according to their behavior in the project we will follow the same approach used by Calefato et al.~\cite{Calefato@ComeBackToContribute}, thus we will calculate for each developer the Truck Factor~\cite{TruckFactors}}.  

\reviewerTwo{Compared with RQ$_1$, when clustering developers based on their median activity break duration, we \textbf{hypothesize} stronger correlations between the antipatterns, code smells, and SonarQube metrics (\Cref{tab:Metrics1} and \Cref{tab:Metrics2}) and activity break duration (H$_{2.1}$). 
The same behavior is expected for the readability metrics (\Cref{tab:Metrics4}) with a stronger inversely proportional correlation with the activity break duration (H$_{2.2}$).}




\subsection{Context}
\label{sec:Context}
\reviewerOne{We will use projects included in available datasets (e.g., Technical Debt Dataset~\cite{LenarduzziPromise2019} version 2.0, Pandora~\cite{Pandora}) that fulfill our criteria:
developed in Java, older than three years, more than 500 commits and 100 classes, and usage of an issue tracking system with at least 100 issues reported.
In  addition to capturing and depicting reality, we are interested in projects that are using SonarCloud in their development process to avoid launching SonarCloud afterward which can lead to inaccurate results because in that case, we would be analyzing problems that the developers would not be aware of.
Finally, we are interested in projects that can be considered mature.
In case the available datasets do not contain the information required we will consider the possibility to extend them or creating a new one.}

\subsection{Verifiability and Replicability}
To allow verifiability and replicability, we will make all the raw data available in our online appendix, including the different scripts we will use in the paper.

\section{Data Collection and Data Analysis}
\label{sec:Methodology}

\subsection{Data collection}
\label{sec:ReadabilityCheckS1} 
\reviewerOne{To answer our RQs, we will find the projects that full fill our criteria and we will collect different software metrics.} In particular, for this analysis, we aim to extract the proxy metrics described in~\Cref{sec:Goal} to estimate, for example, the correlation between the code complexity and the developer’s cognitive perception of the code complexity as previously done by Arisholm \etal~\cite{arisholm2010systematic} for the Line-of-Code (LOC) proxy metric or as in the case of Nagappan and Ball~\cite{nagappan2005use} regarding code churn.
It is worth noticing that some of them could be already included in the selected dataset, while others must be evaluated project-wise. 

\subsection{Calculate project behavior towards code quality}
\label{team-basline}
For each project and for each commit, we will compute the \emph{delta} ($\Delta$) of the aforementioned metrics between that commit and the commit immediately before, to establish whether there was an increase ($\Delta>0$), a decrease ($\Delta<0$), or no variation in the metric values ($\Delta=0$) caused by the actions carried out by the developer. 
The interpretation of the results depends on the specific metrics.

\subsection{Extract activity breaks}

For each developer, we will extract the activity break time (in days) as defined in Section~\ref{sec:Goal}. Days will be grouped along the \emph{last} commit of 
the day. This is justified by the assumption that a user committing several times in a day has not forgotten the code between those commits. Thus, activity breaks will always be
positive natural numbers. 



\subsection{Iteration One}

\subsubsection*{Correlate \reviewerFour{activity breaks with metric values}}
\label{sec:Iteration1Corr}

For each developer, we will select the related commits and, for each metric, we will consider the $\Delta$ computed between that commit and the commit immediately before (which may or may not have been made by the same developer). We will correlate the $\Delta$ values with \reviewerFour{the activity break time}.

In order to account for a non-linear forgetting rate,
as justified by the Ebbinghaus curves presented before,
we will compute \emph{piecewise correlations}, based on a 
piecewise linear regression model \cite{PLR70}.
In a piecewise regression model (also known as \emph{segmented regression}) the independent variable is
partitioned into a given number $n$ of intervals, and a regression model is fit into each of the intervals to clarify its relationship to the dependent variable. We will use a linear regression using the least squares 
method to best fit the data on each of the segments or \emph{bins}.
A fundamental step of piecewise regression is the decision on where to separate the different segments, known as a \emph{breakpoint}. The ideal breakpoint would maximize the difference in slopes between the regression models before and after the breakpoint. There are different strategies for finding such a breakpoint. A fast and robust approach is to group data points with a ``similar'' slope through a clustering method like a decision tree.

We will compute a piecewise regression model to understand the relationship between the activity break duration (independent variable) and the delta for each metric (dependent variable). 
In order to find the best descriptor, we will
test different numbers of bins (from 3 up to a maximum of 10) and different clustering strategies, and will choose the model that presents the smallest error w.r.t.\ the data. In case there are too many data points with the same activity break (i.e., where the dependent variable is the same), we will group them in a representative set using \reviewerOne{centroid-based clustering to limit the cases to a pre-defined number of data points which is coherent with the remaining data. The choice of the number of centroids is made to explicitly regularise outliers in the data.}

Yet, we are not interested in the regression models
\emph{per se}, but rather as a means to understand the impact of forgetting (longer activity breaks) on the quality of the code. An important piece of information is given by the activity breakpoints, which tell us the activity break lengths where the impact on the variable changes behaviors; in other words, they can suggest the \emph{critical} break lengths where the consequences of forgetting the code become more obvious.
To provide an adequate measure of the
influence of the activity break on the metric value, the regression models will be used to compare the differences between the predictions for change in a 1-day break among the available segments. 


\begin{figure}[!t]
    \centering
    \includegraphics[width=\columnwidth, trim={0 0 0 40},clip]{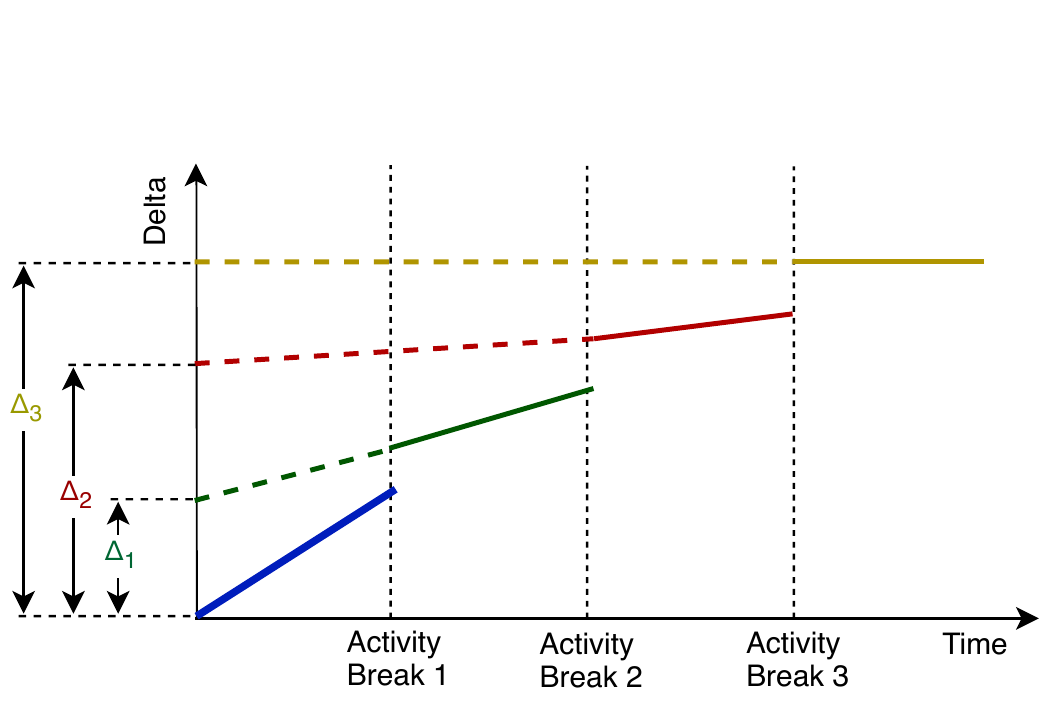}
    \caption{Activity break influence measurement}
    \label{fig:corr}
\end{figure}

Figure~\ref{fig:corr} depicts a dataset where the activity break time (independent variable) is partitioned into four segments, with a linear regression associated with each segment. On the left, we see the difference in the predicted $\Delta$ at 1-day break between the model of the first bin (no forgetting observed) and each of the remaining bins. The breakpoints represent the moments where the behavior w.r.t.\ the break time changes.

After constructing the piecewise regression model and 
computing the differences in the segment behavior as described above, we will make a statistical analysis to verify whether the differences are statistically significant and whether the significance increases as the bin includes longer activity breaks, as our hypothesis suggests. We will also verify the differences between developers, following an inter-study analysis.


\reviewerThree{One can think of several confounding factors that
may  bias the analysis. For instance, the factors shown in Table \ref{tab:Metrics3}---how much of the existing
code was written by the developer, and how much of their previous
work was modified by someone else---may greatly affect the quality
of each commit, but one should not forget that there is no 
an established method for identifying a pre-specified set of important 
confounders and in practice, confounding is not fully overcome 
\cite{Coch22}. To alleviate the effect of these confounders, we will
use a regression detection model and an analysis of covariances
(ANCOVA)~\cite{McNamee500,DAE19}. Moreover, we verify the false positive by a manual inspection that will be done by two authors - and including a third one in case of disagreement. 
}

\begin{table}[!t]
\footnotesize
\centering
\caption{Anti-Patterns and Code Smells collected in this work} 
\label{tab:Metrics1} 
\begin{tabularx}{\columnwidth}{@{}l|X@{}|c@{}}
\hline 
\textbf{Abbrev.} & \textbf{Metric}	& \reviewerTwo{\textbf{H$_{1.1}$}/\textbf{H$_{2.1}$}}	\\	\hline
PCS  &  Code Smells (8) detected by Ptidej: 
Feature envy, Inappropriate intimacy, Large class, Lazy class, Refused bequest, Speculative generality, and Swiss army knife~\cite{Fowler1999} & {\reviewerTwo{\reviewerTwo{$\uparrow$}}}\\\hdashline[1pt/1pt]
PAP & Anti-Patterns (9) detected by Ptidej: Blob, Class data should be private, Downcasting, Excessive use of literals, Functional decomposition, God Class, Orphan variable or constant class, Spaghetti code, and Tradition breaker~\cite{BrownAntipatterns} &  {\reviewerTwo{\reviewerTwo{$\uparrow$}}}
\\\hline
\end{tabularx}
\end{table}

\color{red}
\begin{table}[ht!]
\footnotesize
\caption{SonarQube metrics collected in this work} 
\label{tab:Metrics2} 
\begin{tabularx}{\columnwidth}{@{}l|X@{}|c@{}}

\hline 
\textbf{Abbrev.} & \textbf{Metric}	& \reviewerTwo{\textbf{H$_{1.1}$}/\textbf{H$_{2.1}$}}	\\	\hline
NOC 	&	\# of lines containing either comment or commented-out code	&  {\reviewerTwo{$\uparrow$}}\\\hdashline[1pt/1pt]
NOCD 	&	Density of comment lines\textsuperscript{a}	&  {\reviewerTwo{$\uparrow$}}\\\hdashline[1pt/1pt]
COM	&	Cyclomatic Complexity per function\textsuperscript{b}&  {\reviewerTwo{$\uparrow$}}\\\hdashline[1pt/1pt]
FC 	&	Complexity average per function	&  {\reviewerTwo{$\uparrow$}}\\\hdashline[1pt/1pt]
COGC 	&	Cognitive complexity\textsuperscript{c}	&  {\reviewerTwo{$\uparrow$}}\\\hdashline[1pt/1pt]
DL 	&	\# of lines involved in duplications	&  {\reviewerTwo{\textbf{$\uparrow$}}}\\\hdashline[1pt/1pt]
DB 	&	\# of duplicated blocks of lines	&  {\reviewerTwo{$\uparrow$}}\\\hdashline[1pt/1pt]
DF 	&	\# of files involved in duplications	&  {\reviewerTwo{$\uparrow$}}\\\hdashline[1pt/1pt]
DLD 	&	Density of duplicated lines\textsuperscript{d}&  {\reviewerTwo{$\uparrow$}}\\\hdashline[1pt/1pt]
NTI 	&	\# of all SonarQube issues &  {\reviewerTwo{$\uparrow$}}\\\hdashline[1pt/1pt]
BUG 	&	\# of SonarQube BUG issues	&  {\reviewerTwo{$\uparrow$}}\\\hdashline[1pt/1pt]
CS 	&	\# of SonarQube CODE SMELL issues	&  {\reviewerTwo{$\uparrow$}}\\\hdashline[1pt/1pt]
SV & \# of SonarQube SECURITY VULNERABILITIES issues&  {\reviewerTwo{$\uparrow$}}\\\hdashline[1pt/1pt]
BLOCKER & \# of SonarQube BLOCKER issues&  {\reviewerTwo{$\uparrow$}}\\\hdashline[1pt/1pt]
CRITICAL & \# of SonarQube CRITICAL issues&  {\reviewerTwo{$\uparrow$}}\\\hdashline[1pt/1pt]
MAJOR & \# of SonarQube MAJOR issues&  {\reviewerTwo{$\uparrow$}}\\\hdashline[1pt/1pt]
MINOR & \# of SonarQube MINOR issues&  {\reviewerTwo{$\uparrow$}}\\\hdashline[1pt/1pt]
INFO & \# of SonarQube INFO issues&  {\reviewerTwo{$\uparrow$}}\\\hdashline[1pt/1pt]
TD & Squale index\textsuperscript{e} &  {\reviewerTwo{$\uparrow$}}\\\hdashline[1pt/1pt]
RRE & Reliability remediation effort\textsuperscript{f}&  {\reviewerTwo{$\uparrow$}} \\\hdashline[1pt/1pt]
SRE& Security remediation effort\textsuperscript{g}&  {\reviewerTwo{$\uparrow$}}
\\\hline
\end{tabularx}

\vspace{3pt}
\ff{a}{\# of comment lines $\div$ (\# of lines of code + \# of comment lines) $\times$ 100}
\ff{b}{calculated based on the \# of paths through the code}
\ff{c}{how difficult it is to understand the code based on various criteria like control flow, nesting, or recursion}
\ff{d}{(\# of duplicated lines $\div$ \# of lines) $\times$ 100}
\ff{e}{accumulated technical debt based on issues classified as CODE SMELL}
\ff{f}{accumulated technical debt based on issues classified as BUG}
\ff{g}{accumulated technical debt based on issues classified as SECURITY VULNERABILITY}
\end{table}

\color{black}

\begin{table}[ht!]
\footnotesize
\caption{Redability Metrics collected in this work} 
\label{tab:Metrics4} 
\begin{tabularx}{\columnwidth}{@{}l|X@{}|c@{}}
\hline 
\textbf{Abbrev.} & \textbf{Metric}	&  \reviewerTwo{\textbf{H$_{1.2}$}/\textbf{H$_{2.2}$}}	\\	\hline
CIC & Comments and identifiers consistency 
overlap between the terms used in function comments and the ones in the function bodies &{\reviewerTwo{$\downarrow$}}\\\hdashline[1pt/1pt]
CIC$_{syn}$ & Comments and identifiers consistency, extended considering synonym terms &{\reviewerTwo{$\downarrow$}}\\\hdashline[1pt/1pt]
ITID & Identifier terms in dictionary\textsuperscript{a}&{\reviewerTwo{$\downarrow$}}  \\\hdashline[1pt/1pt]
NMI & Narrow meaning identifiers\textsuperscript{b}&{\reviewerTwo{$\downarrow$}} \\\hdashline[1pt/1pt]
CR & Comments readability&{\reviewerTwo{$\downarrow$}} \\\hdashline[1pt/1pt]
NM & \# of meanings\textsuperscript{c} &{\reviewerTwo{$\uparrow$}}\\\hdashline[1pt/1pt]
TC & Textual coherence\textsuperscript{d}& {\reviewerTwo{$\downarrow$}}\\\hdashline[1pt/1pt]
NOC & \# of concepts\textsuperscript{e} &{\reviewerTwo{$\uparrow$}}\\\hdashline[1pt/1pt]
NOC$_{norm}$ & NOC normalized on the \# of statements& {\reviewerTwo{$\uparrow$}}\\\hline
\end{tabularx}

\vspace{3pt}
\ff{a}{\% of identifiers used in the code that are also part of the English dictionary}
\ff{b}{sum of the particularity of the identifiers}
\ff{c}{polysemy level of the terms appearing in the methods bodies}
\ff{d}{overlap between the terms used in the pairs of syntactic blocks}
\ff{e}{\# of topics detected among statements}
\end{table}

\begin{table}[ht!]

\footnotesize
\centering
\caption{Confounding factors} 
\label{tab:Metrics3} 
\begin{tabular}{@{}l|L{2.5cm}|L{4.5cm}@{}}
\hline 
\textbf{Abbrev.} & \textbf{Metric}	& \textbf{Rationale}	\\
\hline
LMOD  & \% of lines modified by distinct developers between the previous commit and the actual commit & Other developers might have modified the code in the meantime. Mistakes, the introduction of code smells, etc. might not be due to forgetting, but to a misinterpretation of the modifications introduced by others. \\\hdashline[1pt/1pt]
OEXP  &  \% of lines authored in the project up to considered commit & Developers that participated to a large extent to the project (due to their repeated exposure to the code) might forget slower than others.\\\hdashline[1pt/1pt]
\hline
\end{tabular}
\color{black}
\end{table}

\subsection{Iteration Two}
We now describe the step(s) regarding Iteration 2. 

\subsubsection{Classify developers according to their contribution in the project}
Given the information about the activity breaks by each developer, \reviewerOne{as previously done by Calefato et al.~\cite{Calefato@ComeBackToContribute} we will characterize
different developers according to the Truck Factor~\cite{TruckFactors}. 
Specifically, we will analyze the 
%
Truck Factor of each developer, and construct classes depending on their relative behavior 
\emph{average}, \emph{inactive}, and \emph{super inactive}
classes are formed by each of the five quintiles, respectively. That is, super active developers are the 20\% with the lowest average Truck Factor value, and so on.}




\subsubsection{Correlate breaks with metric values for each developer classification} 

Following a strategy akin to Iteration One (Section~\ref{sec:Iteration1Corr}), we will find a correlation between the break
time and the deltas for each metric value. However, in this case, rather than focusing on the specific break time of the developer, we take as the main feature their \reviewerOne{Truck Factor}~\cite{TruckFactors}. In this case, we will analyze the impact of a developer's commitment in each specific metric value, depending on their relative activity in the project. 
To achieve this, we will compute the correlation between the average activity break time (independent variable) and the associated delta (dependent variable). The bins in this case are constructed following the developer classification breaks. The remaining analysis is made through the same strategy described in the previous section.

\section{Execution Plan}
\label{sec:Execution}

We now explain the execution plan we scheduled according to the study design we defined in the previous sections.


\subsection{Data Collection}
\label{execplan:stepa}
\reviewerOne{First of all, we will identify the most suitable projects dataset. We are aware that not all data will be available in the dataset, so we aim to apply the following methodology to calculate process metrics and code readability of projects where source code is developed relying on the \textsc{ versioning system Git} and hosted on a publicly available hub like \textsc{GitHub}.}
First, for each project, we clone the online repository that includes code changes performed on all branches during the software development. Second, we parse the cloned repository with \textsc{PyDriller}~\cite{spadini:2018:pydriller}, a lightweight \textsc{Python} framework designed to ease the mining of \textsc{Git} repositories.
This framework simplifies the retrieval of the two versions of each file changed in a commit, i.e., the version before and after the committed change. By following this approach, we can calculate the process metrics as described in Table \ref{tab:Metrics3}, considering the evolution of the changes applied on each file per developer.
Finally, we will use the  tool developed by Scalabrino et al.~\cite{scalabrino2018ACM} to obtain the values of the readability metrics of these two versions of each file.



\subsection{Calculate project readability and code quality}
\label{execplan:stepc}
For each commit in the dataset, we will perform the following steps:

1) \reviewerFour{query the dataset and group the commits by package}; 

2) \reviewerFour{for each group,} \reviewerOne{query the dataset to get the quality information described in~\Cref{tab:Metrics1} and~\Cref{tab:Metrics2}}; 

3) collect the information described in Table~\ref{tab:Metrics3} and Table~\ref{tab:Metrics4} calculated in the previous step in \Cref{execplan:stepa};

4) calculate the difference for each metric to obtain the $\Delta$ and we will store the commit identifier and the $\Delta$ in a new database.
    
\subsection{Extract activity breaks}
We need to extract the elapsed time between each commit made by each developer. We will perform the following steps: 

1) \reviewerOne{query the dataset to select all the commits performed by the same developer} \reviewerFour{in the same package};

2) \reviewerFour{extract all the activities performed by the same developer in the project};

3) \reviewerFour{create a single list with all the activities and commits and sort them from oldest to newest};

4) calculate the differences in terms of the number of days between each commit and \reviewerFour{activity}; 

5)  add the number of elapsed days to the dataset created in phase three of \Cref{execplan:stepc}.

\subsection{Iteration 1}

In this iteration we first correlate activity breaks with metric values: we will select a number $n$ of bins, and then for each
developer:
1) compute a piecewise regression model w.r.t.\ the dependent variable;
2) for each segment $i$, given by a linear equation $y=a_i x+b_i$ predict the behaviour $\overline{y}_i(1)$ at value $x=1$; and
3) the \emph{impact of forgetting} at bin $i$ is
the difference between the predicted value for bin $i$ and for bin $1$: $\overline{y}_i(1)-\overline{y}_1(1)$.
Subsequently, we will study the impact of confounding factors.
\reviewerThree{An ANCOVA test will be used to understand the impact
of the confounding factors from Table~\ref{tab:Metrics3} to the 
quality of the results.}

\subsection{Iteration 2}
This iteration consists of two steps: first, we classify developers according to their behavior toward commit frequency. \reviewerOne{We will compute Truck Factor~\cite{TruckFactors} for each developer; the average Truck Factor; and calculate the five quintiles. Each developer will be assigned to one of five bins according to the quintile they belong to.} Then, second, we correlate breaks with metric values for each developer classification. For the developers in each quintile, we will compute a 
(classical) linear regression model of the dependent variable ($\Delta$) w.r.t.\ the independent variable (break time). We will analyze these five models for statistical differences.




\section{Threats to Validity}
\label{sec:ThreatsValidity}
In this section, we discuss the threats that might affect the validity of our empirical study, following the structure suggested by Runeson and H{\"o}st~\cite{Runeson2009}.

\textbf{Construct Validity}. 
The planned study uses two main constructs: interruptions during participation in a project and degradation of source code quality. \reviewerThree{Simply observing participation as  source code commits might overestimate disruptions and underestimate participation, since developers may also comment on or review others' code. We, therefore, include other activities, such as pull requests. In addition, we are not able to capture other forms of interactions such as reading code. However, writing code requires a much higher level of "recall" than reading code, so we consider this aspect secondary.}

To study the degradation of source code quality, we plan to use a set of validated measures from the Technical Debt Dataset and calculate additional readability~\cite{scalabrino2018ACM} and source code metrics (Table~\ref{tab:Metrics3}) directly from the source code repository. Despite our efforts to minimize measurement errors, we cannot rule out the possibility of false positives or errors in the measures obtained using these tools.
The selected projects did not use SonarQube or any of the metrics we are calculating during the analysis time frame. As well as for the vast majority of works on Fowler's code smells, developers did not use the rules adopted in the study. 
Our results reflect exactly what developers would obtain using SonarQube out of the box in their project, without customizing the rule-set.

\textbf{Internal Validity}. 
We are aware that static analysis tools detect a non-negligible amount of false positives~\cite{Johnson2013}. However, since we aim at replicating the same conditions that are commonly adopted by practitioners when using the same tools, we will not modify or remove any possible false positives, to accurately reflect the results that developers can obtain by running the same tools. We are aware that we cannot claim  a direct cause-effect relationship between the commit breaks and the selected software metrics, and that the quality of the code  can be influenced by other factors.  We are also aware that pieces of code with different purposes (e.g., classes controlling the business logic) can be more complex than others, and consequently harder to remember. 

\reviewerThree{Last, we are aware that some human factors that can play a relevant role, cannot be measured (e.g., the developer's age, individual skills, etc.).}

\textbf{External Validity.} We will select projects stem from a very large set of application domains, ranging from external libraries, frameworks, and web utilities to large computational infrastructures. The application domain was not an important criterion for the selection of the projects to be analyzed, but in any case, we tried to balance the selection and pick systems from as many contexts as possible. 
Choosing only one or a very small number of application domains, or projects with similar age or size, could have been an indication of the non-generality of our study, as only prediction models from the selected application domain would have been chosen. 
Since we are considering only open-source projects,  we cannot directly speculate on industrial projects. 
Moreover, we only considered Java projects due to the limitation of the used tools (SonarQube provides a different set of Sonar issues for each language) and the results of projects developed on different languages might not be directly comparable.

\section{Related Work}
\label{sec:RelatedWork}

Klammer and Gueldenberg \cite{Klammer2019} performed a systematic literature review about unlearning and forgetting in organizations, distinguish \textit{unlearning} (intentional) from \textit{forgetting} (unintentional), and examine positive and negative consequences of unlearning and forgetting.

Averell and Heathcote~\cite{Averell} study the form of forgetting curves with an experiment that measures different variables over 28 days to observe forgetting and conclude that exponential forgetting curves are the best fit for their participants. 

A study conducted by Kruger et al.~\cite{Kruger2018} surveyed developers of software projects on file familiarity, suggesting that the forgetting curve of Ebbinghaus\cite{ebbinghaus1885,ebbinghaus2013memory} is applicable in software development and that repetitions have a strong relationship with the familiarity of source code.

Fritz et al.~\cite{Fritz} investigated if a programmer’s activity can be used to build a model of what a programmer knows about a code base and through questions to 19 Java developers about files they worked on regularly or recently, they identified a significant correlation between regularly working on a file and familiarity.
LaToza and Myers~\cite{LaToza} gathered over 300 questions asked by programmers while developing software and categorize them, indicating that developers often ask specific questions about a scenario, highlighting the importance of being familiar with the source code.

\reviewerThree{Calefato et al.~\cite{Calefato@ComeBackToContribute} investigated the life-cycle of developers in Open Source projects to delineate a pattern to identify if a developer is abandoning the project or is taking a break. For this purpose, Calefato et al. developed a methodology to identify the break time of developers. In addition, to calculate the risk of degradation of the project, the authors calculated the Truck Factor~\cite{TruckFactors} for each developer.}

To the best of our knowledge, there are no studies that delve deeper into the phenomenon of source code forgetting by examining whether a correlation exists between a change in the quality of source code and a developer activity break. 

\section{Conclusion}
\label{sec:Conclusion}
In this registered report, we aim to examine the correlation between a developer's activity break and various code quality attributes. 
Our goal is to understand if 1)  breaks between activities impact positively or negatively on code quality and 2) if  developer activity profiles (super active,
active, average, inactive, and super inactive) impact on code quality. 

The results of this work will enable researchers and companies to understand if is beneficial to let developers work on multiple projects in parallel or if it is better to have them focus mainly on one project continuously.

\bibliographystyle{IEEEtran}
\bibliography{reference.bib}

\end{document}